\documentclass{JAC2000}
\usepackage{epsfig}

\usepackage{graphicx}

\begin{document}
\def \NN{$N\overline{N}~~$}
\def \DD{$D^0\overline{D^0}~~$}
\def \D0bar{$\overline{D^0}~~$}
\title{
 NEW  RESULTS IN DIFFRACTIVE PHOTOPRODUCTION 
FROM E687 AT FNAL
}
\author{A.~ ZALLO \\
on behalf of the E687 Collaboration
\thanks{co\,authors\,:~
P.L.~Frabetti ({\bf INFN and Bologna});
H.W.K.~Cheung, J.P.~Cumalat, C.~Dallapiccola, J.F.~Ginkel, W.E.~Johns,
M.S.~Nehring, E.W.~Vaandering ({\bf UC Boulder});
J.N.~Butler, S.~Cihangir, I.~Gaines,P.H.~Garbincius, L.~Garren,
S.A.~Gourlay, D.J.~Harding,P.~Kasper,A.~Kreymer,P.~Lebrun,S.~Shukla,
M.~Vittone ({\bf Fermilab}); R.~Baldini-Ferroli,S.~Bianco,F.L.~Fabbri,
S.~Sarwar ({\bf INFN Frascati}); C.~Cawlfield, R.~Culbertson,
R.W.~Gardner, E.~Gottschalk, R.~Greene,K.~Park,A.~Rahimi,
J.~Wiss ({\bf UI Champaign}); G.~Alimonti, G.~Bellini, M.~Boschini,
D.~Brambilla,B.~Caccianiga, L.~Cinquini,M.~DiCorato,P.~Dini,
M.~Giammarchi,P.~Inzani, F.~Leveraro, S.~Malvezzi, D.~Menasce, 
E.~Meroni, L.~Milazzo,L.~Moroni, D.~Pedrini, L.~Perasso, F.~Prelz, 
A.~Sala,S.~Sala,D.~Torretta  ({\bf INFN and Milano});           
D.~Buchholz,D.~Claes, B.~Gobbi,
B.~O'Reilly  ({\bf Northwestern });J.M.~Bishop,N.M.~Cason, C.J.~Kennedy,
G.N.~Kim,T.F.~Lin,D.L.~Puseljic, R.C.~Ruchti, W.D.~Shephard, J.A.~Swiatek,
Z.Y.~Wu  ({\bf Notre Dame});V.~Arena,G.~Boca, G.~Bonomi,C.~Castoldi,
G.~Gianini, M.~Merlo, S.P.~Ratti, C.~Riccardi, L.~Viola,
P.~Vitulo ({\bf INFN and Pavia}); A.M.~Lopez,L.~Mendez,A.~Mirles, E.~Montiel,
D.~Olaya, J.E.~Ramirez, C.~Rivera, Y.~Zhang  ({\bf Mayaguez, Puerto Rico});
J.M.~Link, V.S.~Paolone, P.M.~Yager ({\bf UC DAVIS}); J.R.~Wilson  ({\bf USC 
Columbia});
J.~Cao,M.~Hosack,P.D.~Sheldon ({\bf Vanderbilt});
F.~Davenport ({\bf UNC Asheville}); K.~Cho, K.~Danyo, T.~Handler ({\bf UT
 Knoxville});
B.G.~Cheon, Y.S.~Chung,J.S.~Kang, K.Y.~Kim, K.B.~Lee,S.S.~Myung ({\bf Korea
University, Seoul})
} \\
\it  Laboratori Nazionali di Frascati dell'INFN -  via E.~Fermi 40, Frascati
I-00044  }

\maketitle

\begin{abstract} 
   A narrow dip structure has been observed at 1.9~GeV/c$^2$
  in a study of diffractive photoproduction of
  the $~3\pi^+3\pi^-$ final state performed by the Fermilab experiment E687.
  Preliminary results on new structures in  $~2\pi^+2\pi^-$ final states
  are also given.
\end{abstract}

\section{INTRODUCTION}

\par
The Fermilab Experiment 687 collaboration has collected
a large sample
of high-energy photoproduction
events, recorded with the E687 spectrometer
\cite{Frabetti:1992au}\cite{Frabetti:1993bn}
during the 1990/91
fixed-target runs at the Wideband Photon beamline at Fermilab.
Although the experiment is focussed on charm physics,
a very large sample of diffractively photoproduced light-meson events
was also recorded.
This paper reports on a study\cite{Frabetti:2001ah}
of the diffractive photoproduction of the $3\pi^{+}3\pi^{-}$ final state
and the
observation of a narrow dip in the mass spectrum at 1.9~GeV/c$^2$.
\par\noindent
Preliminary results on diffractive photoproduction of  $~2\pi^+2\pi^-$ final states
are also shown.

\section{E687 SPECTROMETER}

\par
In E687, a forward multiparticle spectrometer is used to
measure the interactions of high-energy photons on a 4-cm-thick Be target.
It is a large-aperture, fixed-target spectrometer with
excellent vertexing, particle identification, and reconstruction
capabilities for photons and $\pi^0$'s. The photon beam is
derived from the Bremsstrahlung of secondary electrons
of  $\approx 300$ GeV endpoint energy, which were produced by the 800
GeV/$c$ Tevatron proton beam. The charged particles emerging from
the target are tracked by a system of twelve planes of silicon microstrip
detectors arranged in three views.~These provide high-resolution separation
of primary (production) and secondary (charm decay or interaction)
vertices.
The momentum of a charged particle
is determined by measuring its deflections in two analysis magnets of
opposite polarity with five stations of multiwire proportional
chambers. Three multicell threshold \v Cerenkov counters are used to
discriminate between pions, kaons, and protons. Photons and neutral pions
are reconstructed by electromagnetic (EM) calorimetry. Hadron calorimetry
and muon detectors
provide triggering and additional particle identification.

\section{EVENT SELECTION}

\begin{figure}[htb]
\centering
\includegraphics*[width=3.0in,height=3.0in]{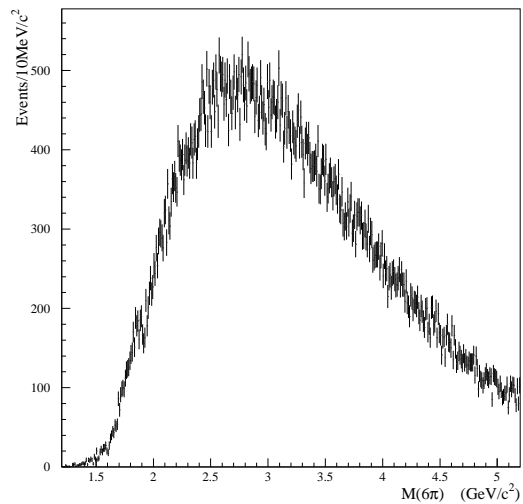}
\caption{ Distribution of $3\pi^+ 3\pi^- $ invariant mass after applying
a cut on the total energy deposited in the calorimeters with
respect to the total energy in the  spectrometer.}
\label{fig:first}
\end{figure}

\par
Pions are produced in photon interactions in the Be target.
The data acquisition trigger requires a minimum energy deposition in the hadron
calorimeters 
located behind the electromagnetic calorimeters and
at least three charged tracks outside the pair region.
The microstrip system and the forward spectrometer
measure the six-pion final state (in this paper,6$\pi$ refers to the $3\pi^+
3\pi^- $ state)
with a mass resolution $\sigma =  10 ~{\rm MeV/c^2}$ at a total
invariant mass of about 2~GeV/c$^2$.
It is required that a single six-prong vertex be reconstructed in the
target region by the microstrip detector, with a good confidence level.
Such a requirement rejects  background due to secondary interactions
in the target.
Exclusive final states are selected by also requiring that the same
number of tracks be reconstructed in the magnetic spectrometer.
The six microstrip tracks and the six spectrometer tracks are required
to be linked, with no ambiguity in associating the microstrip and
spectrometer tracks.
Events with particles identified by the {\v C}erenkov system
as definite electrons, kaons, or protons, or as kaon/proton ambiguous are
eliminated and at least four out of
six particles have to be positively identified as $\pi^{\pm}$.
Particle identification is tested by assuming that one or two out of the
six tracks is a $K^{\pm}$, by computing all two-track invariant mass
combinations, and verifying  that there is no evidence of a peak
at the $K^{*}$ or at the $\phi$ mass.
We eliminated final states with $\pi^0$'s by rejecting events with
visible energy in the electromagnetic calorimeters that was not associated
with the charged tracks. A cut in this variable
($E_{\rm cal}/E_{6\pi} ~\leq~5\%$) is applied
on the calorimetric neutral energy normalized to the
six-pion
energy measured in the spectrometer.
The distribution of the
six-pion invariant mass after these
cuts is shown in Fig.~1. The plot shows a structure
at 1.9~GeV/c$^2$. In the following,
only the 6$\pi$ mass region around this structure will be analyzed.
\par\indent
For diffractive reactions at our energies,
the square of the four-momentum transfer $t$
can be approximated by the square of the total transverse momentum $P_{T}^2$
of the diffractively produced hadronic final state.
Using this definition, the $P_{T}^2$ distribution of
diffractive events, Fig. 2, is well described by two
exponentials:  a coherent contribution with a slope
b$_c = 54\pm 2~$(GeV/c)$^{-2}$
consistent with the Be form factor
 \cite{DeJager:1987qc}
and  an incoherent contribution
with a slope b$_i = 5.10\pm 0.25 ~$(GeV/c)$^{-2}$.

\begin{figure}[htb]
\centering
\includegraphics*[width=3.0in,height=3.0in]{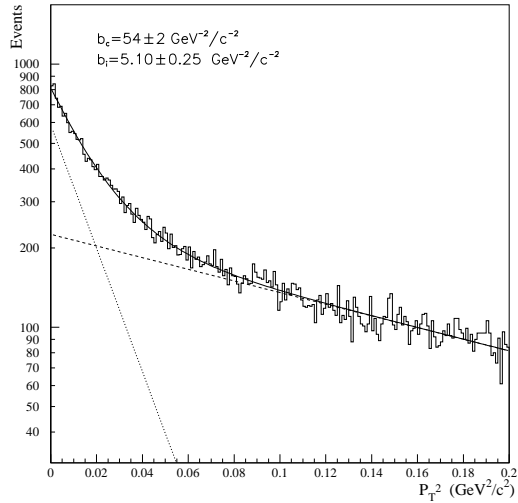}
\caption{ 
Tranverse momentum squared distribution showing the diffractive
peak.
}
\label{fig:second}
\end{figure}

Taking only events with $P_{T}^2 \leq 0.040~ $GeV$^2/c^2$, we evaluated
a contamination of about $ 50 \%$ from nondiffractive events.
This incoherent contribution shows no structure in the
1.2$-$3.0~ GeV/c$^2$ mass range, Fig. 3.
The diffractive mass distribution was obtained by subtracting this
contribution, parametrized by a
polynomial fit, and dividing the yield by the detection efficiency.
\par\indent
The detection efficiency was computed by
modeling diffractive photoproduction of a mass M, using the
experimentally found
slope b$_c$, and simulating the decay M$ \rightarrow{6\pi}$
according to phase space \cite{genbod}.
There is no threshold or discontinuity for the efficiency, particularly in the
region of the dip structure. At $1.9 ~{\rm GeV/c}^2$, the (self-normalized
relative) efficiency {\it A} varied as
{\it dA/A/dM}$_{6\pi}$ = 0.15/$~{\rm GeV/c}^2$.
The efficiency and the efficiency-corrected distribution of
the six-pion invariant mass for
diffractive events,
in the mass range 1.4$- 2.4~{\rm GeV/c}^2$, are shown in Fig.~4.
There  no evidence, albeit with large combinatorial backgrounds, for
resonance substructure, e.g., $\rho^0 \rightarrow \pi^+ \pi^-$, in the 6$\pi$
data below $M_{6\pi} = 2.0~{\rm GeV/c}^2$, either at the mass region of the dip
or in nearby sidebands. Similarly, the efficiency or
acceptance exhibited no threshold, edge, or discontinuity
over the entire mass region observed, when the six pion state
was simulated as a sequence of decays of intermediate two-body
resonances, for example $a_1^+ + a_1^- \rightarrow (\rho^0 \pi^+) + (\rho^0 \pi^
-)
\rightarrow (\pi^+ \pi^- \pi^+) + (\pi^+ \pi^- \pi^-)$, even under extreme
assumptions  of full longitudinal or transverse polarizations
for the initial state.
\par
\begin{figure}[htb]
\centering
\includegraphics*[width=3.0in,height=3.0in]{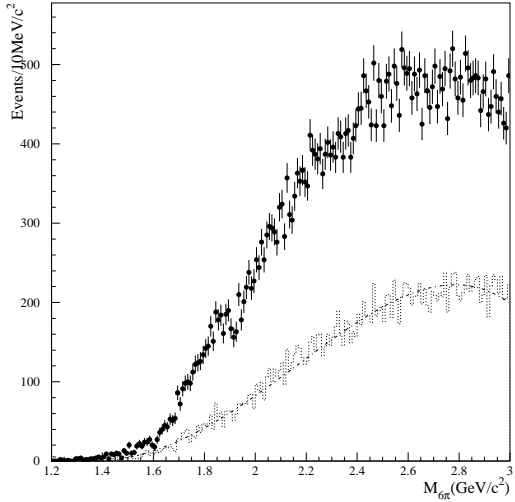}
\caption{ 
 Distribution of $3\pi^+ 3\pi^- $ invariant mass in the
1.2$-$3.0~ GeV/c$^2$ mass
range: coherent plus incoherent contribution.
Dotted distribution: incoherent contribution.
}
\label{fig:third}
\end{figure}

\indent
The presence of a dip at 1.9~GeV/c$^2$ was verified by
several systematics checks.
Diffractive photoproduction of \DD 
pairs or the associated production of  
\D0bar plus a charm baryon at low t 
(where the decay products of the other 
charm meson or charm baryon are missed) followed by the 6 $\pi$ decay
of the D$^0$ or \D0bar
are estimated  by Monte Carlo simulation to be negligible contributions.
It was also checked that demanding more stringent  cuts (i.e., requiring that
all six particles be identified
as $\pi^{\pm}$, applying a sharper cut
on the calorimeter neutral energy, or subtracting the
incoherent contribution bin by bin) increases the statistical
errors without significantly affecting the behavior shown in
Fig.~\ref{fig:third}.
\par\indent
A three-parameter polynomial
fit was performed, solid line in Fig.4,
to explore the hypothesis that any structure in this distribution
is a statistical fluctuation.
The normalized residual distribution,
evaluated for each 10-MeV/c$^2$ bin,
is good in the full invariant mass range,
with the exception of the
interval centered at 1.9~GeV/c$^2$, the region
of the claimed dip, where a poor $\sim~ 10^{-3}$ confidence level
interval, is obtained,
making it highly unlikely that the observed dip is a statistical fluctuation.
Incoherently adding a Breit-Wigner to the fit does not improve the fit
quality much.

\par
\section {FITTING THE SIX-PION INVARIANT MASS DISTRIBUTION}
\begin{figure}[htb]
\centering
\includegraphics*[width=3.0in,height=3.0in]{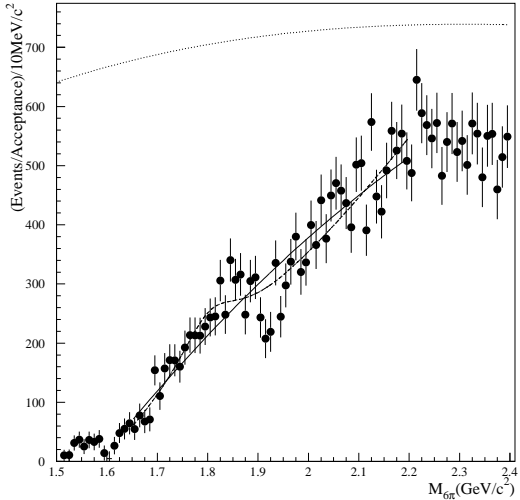}
\caption{ 
Acceptance-corrected distribution of $3\pi^+ 3\pi^- $ invariant mass
for diffractive events after subtracting incoherent contribution.
Solid line: second-order polynomial fit. Dashed line: polynomial fit
with incoherently-added Breit-Wigner.
Upper dot line: relative detection efficiency (arbitrary normalization).
}
\label{fig:fourth}
\end{figure}
\par
\begin{figure}[htb]
\centering
\includegraphics*[width=3.0in,height=3.0in]{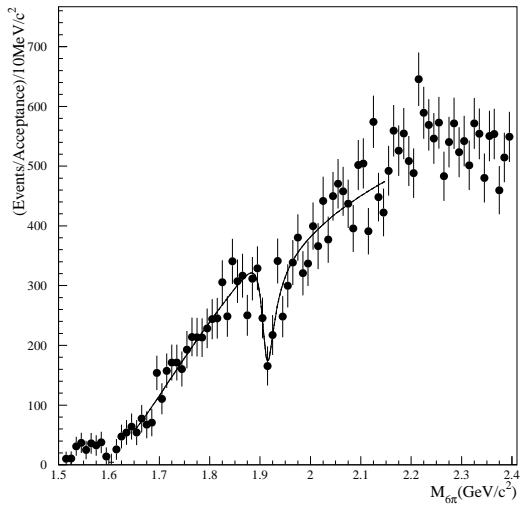}
\caption{ 
 Acceptance-corrected distribution of $3\pi^+ 3\pi^- $ invariant
mass for diffractive  events.
The mass resolution has been unfolded. Fit parameters are listed in 
Table 1.
}
\label{fig:fifth}
\end{figure}

Because of the narrow width, the E687 spectrometer mass
resolution, $\sigma = 10 ~{\rm MeV/c}^2$ at 2 GeV/c$^2$,
was unfolded
by applying the method described in Ref.~\cite{NIM}:
the experimentally observed data distribution $r(x)$ and the unfolded mass
distribution $a(x)$ are related by
$a(x)~\sim~r(x)~-~0.5 \sigma^2 \cdot r(x)^{''}$, where $r(x)^{''}$ is the
second derivative with respect to M$_{6\pi}$ for the observed distribution.
This relationship results from applying a Fourier transform
and approximating the resolution function by
$g(x)~=~ exp(-\frac{\sqrt{2}|x|}{\sigma})$, which is
marginally different from a Gaussian.
A fit similar to the following one is used to obtain $r(x)^{''}$ from the unfold
ed data.
 The data after unfolding are
shown in Fig.~\ref{fig:fifth}.
\par\indent
The dip structure at 1.9~GeV/c$^2$ has been characterized by a two-component
fit, adding coherently a relativistic Breit-Wigner resonance to a
diffractive continuum contribution.
The continuum probability distribution \(F_{JS}(M)\) has been modeled after
a Jacob-Slansky diffractive parameterization\cite{jacob}, plus a constant term
$c_0$

 \[ F_{JS}(M) = f_{JS}^2(M) = c_0 + c_1 \frac{e^{\frac{-\beta}{M-M_0}}}{(M-M_0)^
{2-\alpha}}
  \]
 
The Jacob-Slansky amplitude \(f_{JS}(M)\) is assumed to be the purely
 real  ($\phi_{JS} \equiv 0$) square root of the probability function
 \( F_{JS}(M) \).  For the fit, a
relative phase factor $e^{i\phi}$, independent of mass, and a normalizing
factor $a_r$ multiplied a relativistic Breit-Wigner resonance term, giving
the overall amplitude

\[ A(M) = f_{JS}(M) + a_r \frac{-M_r \Gamma e^{i\phi}}{M^2-M_r^2 +
 i M_r\Gamma} \]

Fit results are shown in Table~\ref{tab:unfolded} and in
Fig.~\ref{fig:fifth} for a fitted mass range
from 1.65 to 2.15 {\rm GeV/c}$^2$,
symmetric with respect to the dip. Quantities shown are
the mass and width of the resonance, the amplitude ratio
 $a_{\rm r}/f_{\rm{JS}}(M_{\rm r})$  between the Breit-Wigner function and
 the Jacob-Slansky continuum, the
relative phase and the $\chi^2$/dof.
Fit values show consistent evidence for a narrow resonance at
M$_r$ ~=~1.911 $\pm$ 0.004 $\pm$ 0.001 ~GeV/c$^2$ with a width
$\Gamma$ = 29 $\pm$ 11 $\pm$ 4 ~MeV/c$^2$,
where the errors quoted are statistical and systematic,
respectively.
The fit values shown in Fig.~\ref{fig:fifth} and represented by the parameters
of Table 1 are stable with acceptable $\chi^2/{\rm dof}$ over varying mass
ranges from 1.65 to 2.3 {\rm GeV/c}$^2$.
We quote as systematic error the sample variance of the fit values
due to our choice of fit mass range.
The  quality of the fit deteriorates somewhat as the upper limit
of the fit for this simple model is extended from
2.1 to $3.0~{\rm GeV/c^2}$.
However, the only fit parameter that is affected is the
width, which varies from $29\pm 11 \, {\rm MeV/c}^2$ to
$40 \pm 20 \, {\rm MeV/c}^2$.
\par
\begin{table}
\caption{Fit results for a mass range from 1.65 to 2.15 $GeV/c^2$
\label{tab:unfolded}
}
\begin{center}
\begin{tabular} {|l|r|l|r|}
\hline\hline
M$_r$ (GeV/c$^2$) & $1.911 \pm 0.004$ \\
$\Gamma$ (MeV/c$^2$) & $29 \pm 11$ \\
$a_{\rm r}/f_{\rm{JS}}(M_{\rm r})$ & $0.31 \pm 0.07$ \\
$\phi$ (deg.) & $62 \pm 12$ \\
$\chi^2$/dof & 1.1 \\
\hline\hline
\end{tabular}
\vfill
\end{center}
\end{table}
\par
\par
\section{DIFFRACTIVE PHOTOPRODUCTION OF 2($\pi^+\pi^-$) EVENTS}
\par

About one million of 2($\pi^+\pi^-$) events 
have been recorded  by E687 in the 1990/91 fixed target runs
at the Fermi National Accelerator Laboratory.Their analysis is in progress and in the following
only a  preliminary presentation is given.
As in the $3\pi^+ 3\pi^- $ study, exclusive
 final states are selected based on the total number of charged tracks seen in
 the spectrometer and final states with $\pi^0$'s are rejected by requiring 
no visible energy in the electromagnetic calorimeters.
The particle identification has been done applying the same cuts described in
section 3. The t-distribution shows
a  $P_{T}^2$ slope of about 60 (GeV/c)$^{-2}$ , in good agrement with
the $3\pi^+ 3\pi^- $ sample.
Taking only events with $P_{T}^2 \leq 0.0625~ $GeV$^2/c^2$, our 2($\pi^+\pi^-$)
 sample remains with a contamination of about $ 25 \%$ from nondiffractive events.
\par
\begin{figure}[htb]
\centering
\includegraphics*[width=3.0in,height=5.0in]{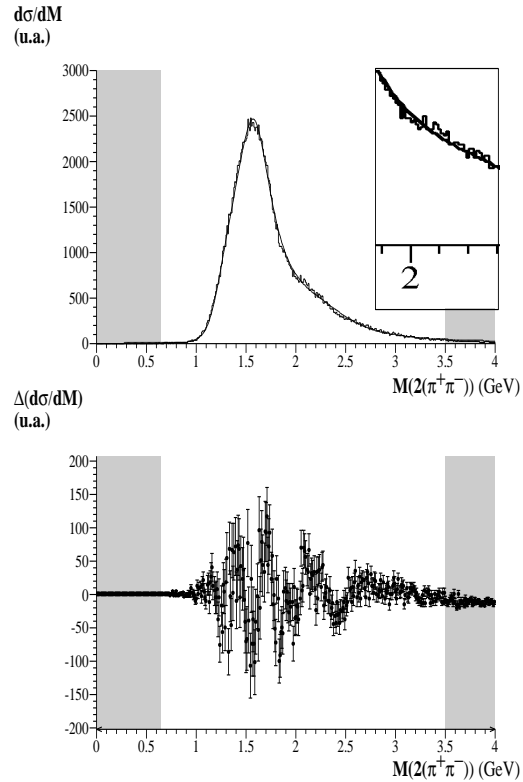}
\caption{ 
 Acceptance corrected 2($\pi^+\pi^-$) mass  (top) and
residual (data - fitted values) distributions (bottom).The solid line is
 the 3 P-Wave Breit-Wigner fit.
}
\label{fig:sixth}
\end{figure}

\begin{figure}[htb]
\centering
\includegraphics*[width=3.0in,height=5.0in]{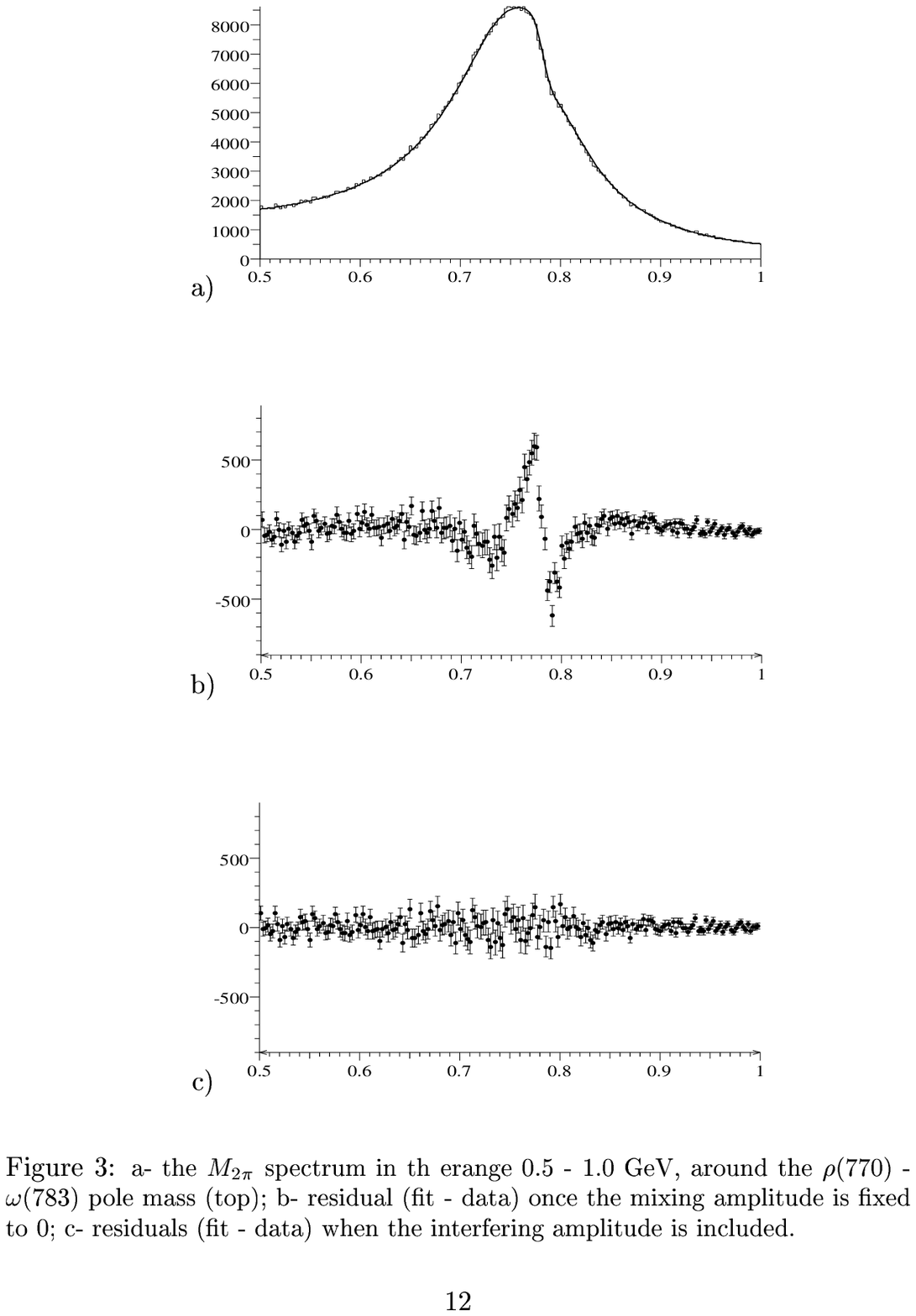}
\caption{ 
 a- the M$_2\pi$ spectrum in the range 0.5-1.0 GeV/c$^2$, around the 
$\rho(770)- \omega(783)$ pole mass (top); b- residual (fit-data) once 
the mixing amplitude is fixed to 0; c- residuals (fit-data) when the
interfering amplitude is included.
}
\label{fig:seventh}
\end{figure}

\indent
 The acceptance corrected 2($\pi^+\pi^-$) mass  spectrum is shown in 
Fig.~\ref{fig:sixth}(top).Although we expect, to first order, 
that the bump in the mass spectrum below 2.0 GeV/c$^2$ is dominated by
the $\rho$(1450) vector meson, our fit is not  good (solid line)  even when
we add 3 P-Wave  interfering Breit-Wigner: the $\chi^2$ is 467.0 for 274 
degrees of freedom.As the
acceptance is almost flat for M$_{4\pi} \sim~ 1.6~  $GeV$^2/c^2$,
the deviations observed should be of physical origin.
The residuals (data - fitted values)  shown in Fig.~\ref{fig:sixth}(bottom)
clearly show  structures with a non negligible statistical significance.
Their study is in progress. For the time being we can say that the ratio of
the residuals to the total yield in 4$\pi$ is of the same order of the 
$\rho-\omega$ mixing effect in 2$\pi$\cite{2pi},shown in 
Fig.~\ref{fig:seventh}.
\section{CONCLUSIONS}
The diffractive photoproduction of  $3\pi^{+}3\pi^{-}$
 has been studied by E687. Evidence has been found for a narrow structure
near $M_{6\pi}$ = 1.9 GeV/$c^2$.  If this dip is characterized as the
destructive interference of a resonance with the continuum background, then
the parameters of this resonance would be
M$_r$ ~=~1.911 $\pm$ 0.004 $\pm$ 0.001 ~GeV/c$^2$,
with $\Gamma$ = 29 $\pm$ 11 $\pm$ 4 ~MeV/c$^2$.Such a resonance could be 
assigned the photon quantum numbers (J$^{\rm PC}=1^{--}$) and G=+1, I=1 due
to the final state multiplicity. 
\par\indent
The possible presence of other  structures  in the 2($\pi^+\pi^-$) final state 
needs further study.

\end{document}